\documentclass{article}

\usepackage{PRIMEarxiv}

\usepackage[utf8]{inputenc} 
\usepackage[T1]{fontenc}    
\usepackage{hyperref}       
\usepackage{url}            
\usepackage{booktabs}       
\usepackage{amsfonts}       
\usepackage{nicefrac}       
\usepackage{microtype}      
\usepackage{lipsum}
\usepackage{fancyhdr}       
\usepackage{graphicx}       
\graphicspath{{media/}}     
\usepackage{subcaption} 
\usepackage{amsmath}

\title{QAGT-MLP: An Attention-Based Graph Transformer for Small and Large-Scale Quantum Error Mitigation
}

\author{
  Seyed Mohamad Ali Tousi \\
  Vision-Guided and Intelligent Robotics Lab (ViGIR) \\
  University of Missouri \\
  Columbia, US\\
  \texttt{STousi@missouri.edu} \\
   \And
  G. N. DeSouza \\
  Vision-Guided and Intelligent Robotics Lab (ViGIR) \\
  University of Missouri \\
  Columbia, US\\
  \texttt{DeSouzaG@missouri.edu} \\
}

\begin{document}
\maketitle

\begin{abstract}

Noisy quantum devices demand error-mitigation techniques to be accurate yet simple and efficient in terms of number of shots and processing time. Many established approaches (e.g., extrapolation and quasi-probability cancellation) impose substantial execution or calibration overheads, while existing learning-based methods have difficulty scaling to large and deep circuits. In this research, we introduce \textbf{QAGT-MLP}: an attention-based graph transformer tailored for small- \textbf{and} large-scale quantum error mitigation (QEM). QAGT-MLP encodes each quantum circuit as a graph whose nodes represent gate instances and whose edges capture qubit connectivity and causal adjacency. A dual-path attention module extracts features around measured qubits at two scales or contexts: 1) graph-wide global structural context; and 2) fine-grained local lightcone context. These learned representations are concatenated with circuit-level descriptor features and the circuit noisy expected values, then they are passed to a lightweight MLP to predict the noise-mitigated values. On large-scale 100-qubit Trotterized 1D Transverse-Field Ising Models -- TFIM  circuits -- the proposed QAGT-MLP outperformed state-of-the-art learning baselines in terms of mean error and error variability, demonstrating strong validity and applicability in real-world QEM scenarios under matched shot budgets. By using attention to fuse global structures with local lightcone neighborhoods, QAGT-MLP achieves high mitigation quality without the increasing noise scaling or resource demand required by classical QEM pipelines, while still offering a scalable and practical path to QEM in modern and future quantum workloads.
\end{abstract}


\section{Introduction}

Quantum Computers are offering polynomial and exponential speedup advantages over their classic counterparts. However, this promise is overshadowed by the inherent noise of physical quantum computers. Such noise limits any practical deployment of quantum computers in real-world scenarios. At the same time, full error correction is computationally expensive (if not impossible). So, practical and accurate Quantum Error Mitigation (QEM) techniques must facilitate their use in increasingly more realistic problems. 

Most state-of-the-art QEM techniques rely on classical statistical methods \cite{li2017efficient, liao2024machine, temme2017error, tsubouchi2023universal}, which, despite their recent achievements, still suffer from large computational overheads (e.g. Zero Noise Extrapolation \cite{li2017efficient}). In the meantime, traditional Machine Learning (ML) methods, such as in \cite{liao2024machine}, lack the full exploitation and awareness of the structures of quantum circuits. In that sense, despite the advancements in transformer models and the introduction of Graph Transformer models \cite{yun2019graph}, very few research have reported the use of attention-based mechanisms to help addressing QEM problems. Perhaps the only exception is found in \cite{bao2025beyond}, where a sophisticated non-message-passing graph transformer model was proposed in this context. However, this approach fails to compare with state-of-the-art machine learning methods such as the Random Forests methodintroduced in \cite{liao2024machine}). Moreover, the excessive computational cost associated with their complex graph transformer model renders their solution practically unfeasible. 

In this research, we maintain that for any practical and accurate ML-based QEM technique to exist, one must resort to a light-weight graph attention mechanism with proper feature extraction. In addition, we propose to encode each quantum circuit as a graph, and to extract two sets of features for each desired qubit: 1) a global graph features, and 2) a local lightcone features. We demonstrate that using these two sets of features, derived from a simple 3-layered graph transformer networks, our model can capture a wide range (i.e. scale) of meaningful structural information from extremely complex quantum circuits (e.g. with over 100 qubits). Our results show a strong improvement over the state-of-the-art methods in both mean and standard deviation of the error. 

To summarize, our contributions are:

\begin{enumerate}
    \item We propose a simple graph transformer architecture for QEM that despite its simplicity, shows powerful performances in QEM scenarios, especially in extremely complex quantum circuits with over 100 qubits. 

    \item We propose to extract two sets of features for each quantum circuits: global and local lightcone features. These features are then proved to essential in having an accurate and reliable QEM method. 

    \item We benchmark our results against the state-of-the-art machine learning methods introduced in literature. We also provide an ablation study on each part of our proposed architecture, showcasing the effectiveness of each part of the pipeline. 
\end{enumerate}
\section{Background and Related Works}


Quantum computing is entering a new phase of maturity in which practical applications are increasingly achievable, marking a shift from theoretical promise to tangible impact~\cite{kim2023evidence, google2025quantum, king2025beyond, chow2024quantum, guenot2025can, soller2025year}. Rapid progress in both hardware and algorithmic development has accelerated the deployment of quantum techniques in real-world scenarios, supported by improvements in qubit coherence, circuit fidelity, and hybrid quantum--classical workflows.

Recent advances have demonstrated the potential of quantum computing across diverse scientific and engineering disciplines. In chemistry and materials science, quantum algorithms have been successfully used for molecular energy estimation, catalyst design, and reaction pathway discovery~\cite{mcardle2020quantum, kandala2017hardware, cerezo2021variational}. In finance and operations research, quantum optimization and sampling techniques such as the Quantum Approximate Optimization Algorithm (QAOA) and quantum annealing have been explored for portfolio optimization and logistics scheduling~\cite{egger2020quantum, orus2019quantum}. In biomedical and pharmaceutical research, quantum-enhanced simulations and hybrid machine learning methods have shown promise for drug discovery and protein folding~\cite{ghazi2025quantum, li2024hybrid, kumar2025towards, jun2025quantum, fairburn2025applications}. These efforts collectively demonstrate how quantum methods can provide new computational advantages for modeling, inference, and optimization in complex, high-dimensional systems.

At the algorithmic level, many of these applications share a common challenge: efficiently solving large-scale, non-convex, or combinatorial optimization problems that are intractable for classical methods~\cite{biamonte2017quantum, farhi2014quantum, peruzzo2014variational}. Quantum algorithms---through superposition, interference, and entanglement---can explore exponentially large search spaces and approximate globally optimal solutions more effectively than their classical counterparts. This unique capability has positioned quantum computing as a transformative paradigm for computational science, optimization, and data-driven modeling in the near term.

Despite these advances, the practical deployment of quantum algorithms remains fundamentally limited by noise and decoherence in near-term quantum hardware. Current devices, often referred to as Noisy Intermediate-Scale Quantum (NISQ) systems~\cite{preskill2018quantum}, operate without full fault tolerance and thus require specialized techniques to ensure computational reliability. In this regime, \textit{Quantum Error Mitigation} (QEM) has emerged as a crucial strategy for improving result fidelity without the heavy overhead of quantum error correction. QEM methods---including extrapolation-based approaches~\cite{temme2017error, li2017efficient}, probabilistic error cancellation~\cite{endo2018practical}, and symmetry verification~\cite{bonet2018low}---aim to infer or reconstruct ideal expectation values from noisy circuit outcomes. These techniques enable meaningful use of NISQ devices for scientific applications by reducing noise-induced biases in observables, and serve as an essential bridge toward fault-tolerant quantum computation. Building upon this foundation, recent research has explored leveraging data-driven and machine learning approaches~\cite{lowe2021unified, liao2024machine} to enhance QEM scalability and adaptivity, motivating our investigation into more expressive, structure-aware models for large-scale quantum circuits.


The idea behind the extrapolation-based QEM methods is to estimate observables at several amplified noise levels and extrapolate to zero noise \cite{temme2017error, li2017efficient}. The Zero-Noise-Extrapolation (ZNE) was also applied to the superconducting quantum computers for chemistry and magnetism tasks \cite{kandala2018extending}. These approaches, however, require multiple noise-scaled executions per circuit which impose a huge overhead computation burden that is not cost and time-efficient. 

Probabilistic Error Cancellation (PEC) was proposed by \cite{endo2018practical} which was an improvement in terms of the overhead quantum computation by representing noisy channel inverse via a quasi-probability mixtures and sampling circuits accordingly to unbiasedly canceling the noise. However, its shot complexity explodes with the sampling overhead, scaling roughly exponentially in circuit depth. It also requires noise characterization. 

Authors of \cite{bonet2018low} proposed a low-cost error mitigation approach by measuring or inferring conserved symmetries in quantum circuits and discard outcomes that violate them. Although this approach does not impose any computational overhead to the process, it can result in a high \textit{shot wastage}, especially in high-noise devices. It also requires extra measurements and experiments to find the symmetries. This shot wastage probability renders the approach being not practical in dealing with large-scale quantum circuits.

\subsection{Machine Learning for QEM}

Data-driven QEM approach was introduced by the authors of \cite{lowe2021unified} in which they generated a large number of simulable Clifford or near-Clifford quantum circuits and fitted a regressor on the noisy to exact expected value. However, their approach requires classical solvability of the training set. This requirement renders the approach not applicable (or at least not optimal) for large-scale non-Clifford circuits (as it's been shown in \cite{liao2024machine} that linear regression is not the best machine learning model to mimic ZNE for large-scale quantum circuits).  

A recent significant contribution to QEM area \cite{liao2024machine} systematically studied the performance of various ML models for error mitigation. In that study, the authors benchmarked a wide range of models – from simple linear regression to more complex neural networks – under realistic conditions. They considered diverse circuit classes, including random circuits and Trotterized Ising model dynamics, and evaluated the methods on both simulated noise and real hardware experiments. As the authors argue in \cite{liao2024machine}, Random Forests were the most successful model in ML-QEM tasks, consistently showing superior performance across all the other machine learning models. However, we argue that using graph encoding of the quantum circuits, especially in large-scale circuits, would provide more insightful features for a Graph Attention model and can result in lower error rates. 

\subsection{Graph Attention Models}

Graph neural networks (GNNs) based on message passing (MPNNs) propagate information along edges via neighborhood aggregation and have become a standard for graph learning \cite{gilmer2017neural}, but they can suffer from limitations such as over-smoothing, and limited long-range interaction. Attention mechanisms partially addressed these issues in Graph Attention Networks (GAT), which apply masked self-attention over local neighborhoods \cite{velivckovic2017graph}.
\section{Methodology}

\begin{figure}
    \centering
    \includegraphics[width=\linewidth]{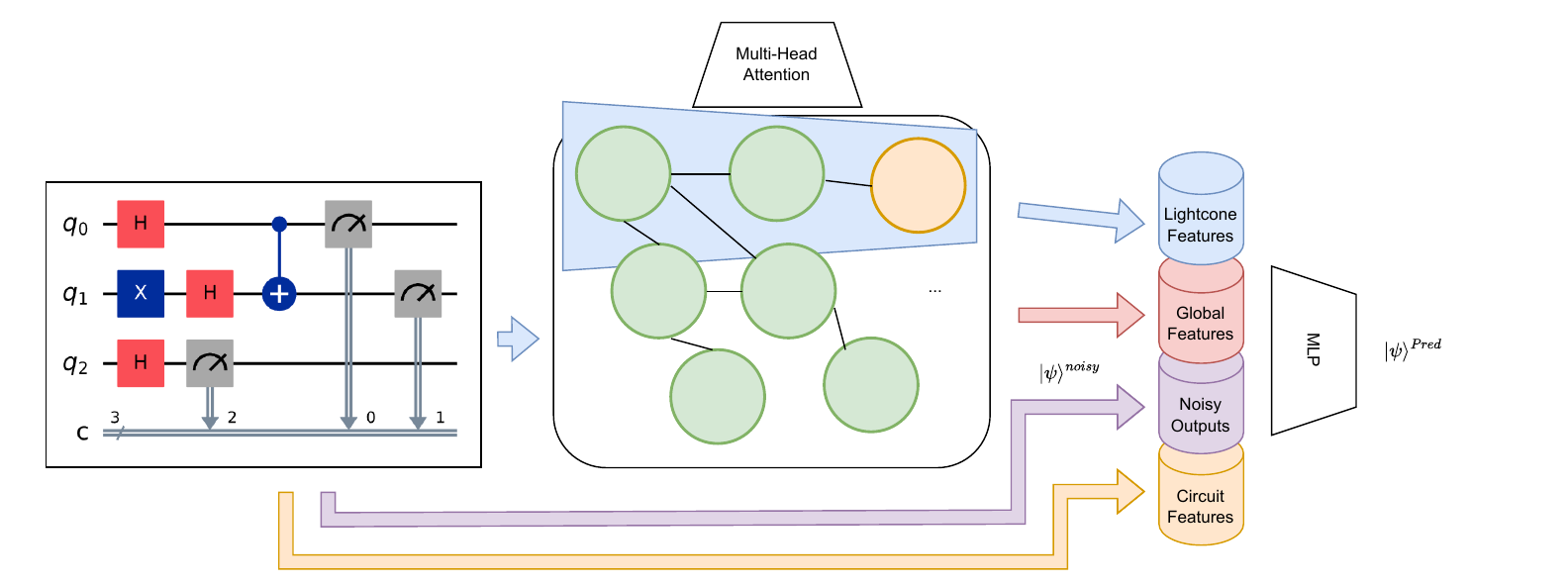}
    \caption{The proposed QAGT-MLP model. The circuit measurement-based and structural features are concatenated with the local lightcone and global circuit-wide contexts to form an input feature map to the MLP. The MLP produces the final estimation of the corrected expected values.}
    \label{fig:pipeline}
\end{figure}

In this section, we describe the building blocks of our proposed QAGT-MLP. We focus on the construction of input features from quantum circuits, the design of our proposed QAGT-MLP architecture, and the training procedure.
Figure \ref{fig:pipeline} shows a schematic of the proposed QAGT-MLP model. 

\subsection{Feature Encoding}
To represent each quantum circuit in a form suitable for machine learning, we adopt a feature encoding strategy introduced by \cite{liao2024machine}, with both local and global descriptors (delineated with purple and orange color coding in Figure \ref{fig:pipeline}):

\begin{itemize}

    \item \textbf{Gate Counts:} For a fixed gate set $\{ \texttt{ecr}, \texttt{sx}, \texttt{x}, \texttt{id}, \texttt{rz} \}$, we compute normalized counts of each gate type across the circuit.
    
    \item \textbf{Angle Binning:} Single-qubit rotation angles are quantized into discrete bins of size $0.025\pi$, producing a histogram that captures the distribution of rotation magnitudes.
    
    \item \textbf{Expectation Values:} The noisy expectation values of the observables are appended directly to the input representation.
    
    \item \textbf{Measurement Bases:} Where applicable, the chosen measurement basis vectors are included as additional binary features.
    
\end{itemize}

The resulting input is a concatenated vector capturing both structural (gate counts and angle binning) and measurement-specific information (noisy expected values and measurement bases). 

\subsection{QAGT-MLP Model}

The second part of the proposed QAGT-MLP is a lightweight Graph-Transformer Model that combines graph-inspired attention with multi-layer perceptron (MLP) layers:

\begin{itemize}

    \item \textbf{Graph Attention Layer:} The first stage consists of an attention mechanism applied over the circuit graph, where qubits and gates and their connectivity form the nodes and edges respectively. Two feature masks are used: 1) a \emph{lightcone mask} (blue in Figure \ref{fig:pipeline}), restricting attention to nodes within the causal cone of a measurement, and 2) a \emph{global mask} (red in Figure \ref{fig:pipeline}), allowing every node to communicate with all others. This hybrid approach captures both local and global circuit-wide context.
    
    \item \textbf{MLP:} The outputs of the attention layer are aggregated and passed through an MLP. The MLP refine the learned representation and estimates the corrected expectation value.

\end{itemize}

Compared to general graph neural networks, QAGT-MLP is extremely shallow and specialized, enabling it to generalize well with limited training data.

\section{Experiments and Results}

\subsection{Dataset}
We use the dataset introduced in \cite{liao2024machine}, which consists of real hardware runs of 100-qubit Trotterized dynamics of the 1D transverse-field Ising model (TFIM), executed on the \texttt{ibm\_brisbane} device. Each circuit is associated with 1) the noisy expectation values obtained from real hardware, and 2) the reference values derived via Zero-Noise Extrapolation (ZNE), which we use as our ground truth labels.

\begin{figure}
    \centering
    \includegraphics[width=\linewidth]{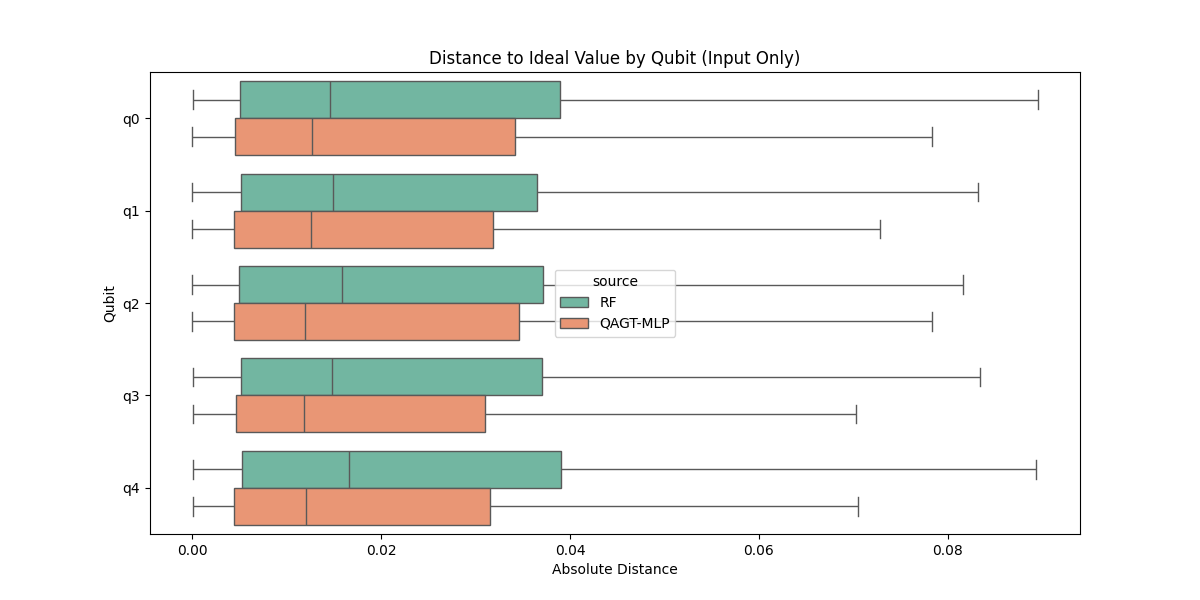}
    \caption{The comparison between the mean absolute error (distance) to the ideal values for each qubit resulted from QAGT-MLP and RF. The QAGT-MLP outperforms RF is all the qubits in both mean distance and its standard deviation.}
    \label{fig:rfvsgt}
\end{figure}

\subsection{Training Procedure}
We train the model in a supervised manner with the following setup:
\begin{itemize}
    \item \textbf{Loss Function:} Mean squared error (MSE) between the predicted and target (ZNE) expectation values.
    \item \textbf{Optimizer:} Adam optimizer with an initial learning rate tuned via grid search.
    \item \textbf{Data Split:} Circuits are divided into 100 circuits for training and 400 circuits for validation. Validation results are used for early stopping and hyperparameter selection.
    \item \textbf{Baselines:} For comparison, we also train the best QEM model introduced by \cite{liao2024machine} which was Random Forests.
\end{itemize}

\subsection{Ablation Studies}

To evaluate the contribution of each component, we perform ablation experiments by disabling the global attention, removing the lightcone mask, or replacing the graph attention with a plain MLP. These variants allow us to quantify the importance of local and global contexts, and attention mechanisms in the performance of QAGT-MLP.

\subsection{Results}

\subsubsection{Overall Performance}

Figure \ref{fig:rfvsgt} shows the comparison between the mean absolute error for each measured qubit derived from 1) the proposed QAGT-MLP and 2) the state-of-the-art random forests models. The proposed QAGT-MLP outperforms RFs in both mean error and its standard deviation for all the qubits.

\subsubsection{Ablation Results}

\begin{figure}[htbp]
    \centering
    \begin{subfigure}{0.45\textwidth}
        \centering
        \includegraphics[width=\linewidth]{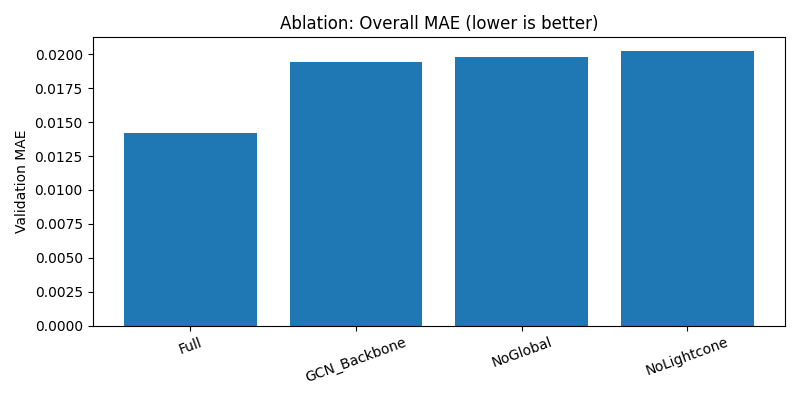}
        \caption{Total mean absolute error.}
        \label{fig:ablation-a}
    \end{subfigure}
    \hfill
    \begin{subfigure}{0.45\textwidth}
        \centering
        \includegraphics[width=\linewidth]{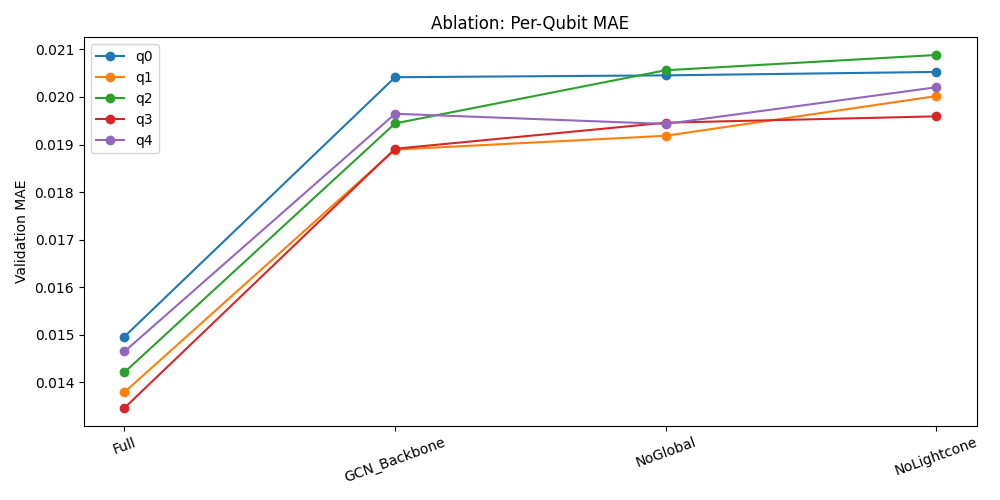}
        \caption{Ablation comparison for each qubit.}
        \label{fig:ablation-b}
    \end{subfigure}

    \caption{The results of ablation study on the proposed components of the QAGT-MLP. The columns are representing: \textbf{Full}: the full proposed QAGT-MLP architecture, \textbf{GCN\_Backbone}: using a graph convolutional network as the backbone instead of Graph Transformers, \textbf{NoGlobal}: not extracting the global context features, and \textbf{NoLightcone}: not extracting the causal lightcone features for each measured qubit.}
    \label{fig:ablation}
\end{figure}

To demonstrate the importance of each proposed QAGT-MLP components, we have presented an ablation study in Figure \ref{fig:ablation} in which \ref{fig:ablation-a} shows the overall error due to removal of each component in the pipeline, and \ref{fig:ablation-b} shows the impact of those removal on the error associated with each qubit. 

\subsubsection{Mimicking ZNE Through Different Trotter Steps}

\begin{figure}
    \centering
    \includegraphics[width=0.9\linewidth]{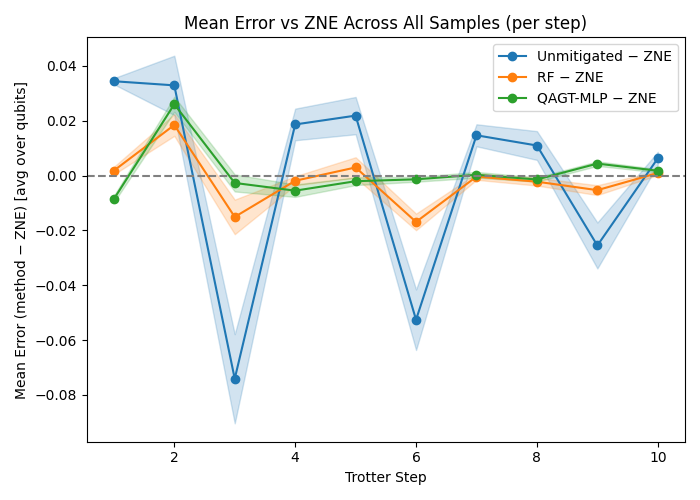}
    \caption{Mean error of the QAGT-MLP, RF, and unmitigated outputs with respect to the ZNE across Trotter steps. The QAGT-MLP exhibits minimal deviation from ZNE, indicating that it can effectively replace ZNE without additional sampling overhead.}
    \label{fig:trotter}
\end{figure}

To evaluate how closely the proposed QAGT-MLP can replicate the ZNE-mitigated results across different Trotter steps, we computed the mean error of each method with respect to the ZNE baseline. Figure \ref{fig:trotter} illustrates the average deviation of the QAGT-MLP, RF, and unmitigated circuit outputs from the ZNE values at each step. As observed, the QAGT-MLP consistently achieves near-zero mean error across all Trotter steps, demonstrating its ability to reproduce ZNE-level accuracy without incurring the additional computational overhead of running ZNE.

\section{Discussion}

In this section, we provide some discussion about the performance of QAGT-MLP and its applicability into real-world quantum error mitigation scenarios. 

\subsection{Model Size}

The proposed QAGT-MLP is an extremely light-weight neural network model with a powerful generalizability across quantum circuit error mitigation. An important point about graph transformers is that the multi-head attention modules in graph transformers have the complexity of $O(N^2)$ in which $N$ is the depth of the graph. This complexity raises the concern of parameter exploding with large graph inputs. However, given its small number of layers and light-weight MLP head, the proposed QAGT-MLP has only 251,393 parameters. This relative small number of parameters enables QAGT-MLP to be applied onto training scenarios with limited or even no state-of-the-art computational power. In fact, all the experiments presented in this study have been reproduced using only CPU power to train the models. 

The other benefit of having this small number of parameters is in the deployment phase of QAGT-MLP. While the proposed model shows superior performance compared to the state-of-the-art machine learning approaches proposed in \cite{liao2024machine}, it is fast and deployable in almost any computational infrastructure. 

\subsection{Computational Costs of ZNE vs. QAGT-MLP}

ZNE is computationally expensive. Given the current scarcity of the real quantum computers and the financial costs associated with running circuits on them, having to run a given quantum circuit multiple times with different noise profiles through the real computers is a huge over-head burden. QAGT-MLP is capable of being trained on the ZNE values once (even with limited data, as in the presented experiments) and deployed in the real-world QEM scenarios.

\begin{table}[h]
\centering
\caption{Comparison of total circuit executions (runtime cost) for ZNE and QAGT-MLP mimicry under different noise factors $m$.}
\label{tab:runtime_cost}
\begin{tabular}{lccc}
\toprule
\textbf{Method} & \textbf{$m$ (Noise Factors)} & \textbf{Total Executions} & \textbf{Relative Cost (×)} \\
\midrule
ZNE (Traditional QEM) & 2 & $2 \times 400 = 800$ & 1.0 \\
ZNE (Traditional QEM) & 3 & $3 \times 400 = 1200$ & 1.0 \\
QAGT-MLP (ours) & 2 & $2 \times 100 + 400 = 600$ & \textbf{0.75×} \\
QAGT-MLP (ours) & 3 & $3 \times 100 + 400 = 700$ & \textbf{0.58×} \\
\bottomrule
\end{tabular}
\end{table}

Assuming that the mimicked quantum error mitigation (QEM) process requires a total of $m$ executions—either of the
mitigation circuits or the circuit of interest itself (for example, digital or analog ZNE typically employs $m=2$ or $3$
noise levels)—the overall runtime cost of the traditional ZNE procedure can be expressed as $m \, n_{\text{test}}$.
In contrast, the total runtime cost of the proposed ML-based mimicry includes both training and inference, resulting in
a total cost of $m \, n_{\text{train}} + n_{\text{test}}$.  
By setting these two costs equal, we derive the break-even ratio between the training and testing circuit counts at
which the computational cost of our mimicry matches that of conventional QEM:
\[
\frac{n_{\text{train}}}{n_{\text{test}}} = \frac{m - 1}{m}.
\]
The mimicry demonstrates a net efficiency advantage when the actual train–test ratio is smaller than $(m - 1)/m$ \cite{liao2024machine}.
Table~\ref{tab:runtime_cost} reports the total circuit execution requirements for both approaches under different
values of $m$, assuming $n_{\text{train}} = 100$ and $n_{\text{test}} = 400$.

\subsection{Locality of The Lightcone Masks}

One of the contributions of the paper is to use features derived by a causal lightcone mask for each measured qubit. Looking at the locality of these masks may provide valuable insights about the data and the types of circuits that we are mitigating. Table \ref{tab:lightcone_locality} shows the average locality metrics among the training data. For a circuit graph with $N$ operation-nodes and a lightcone mask $\mathcal{L}_m \subseteq V$ associated with measured qubit $m$, we define:

\begin{itemize}
    \item \textbf{Coverage:}
    \[
    \text{coverage}_m = \frac{|\mathcal{L}_m|}{N}
    \]
    Fraction of all nodes that belong to the lightcone of qubit $m$. Smaller values indicate stronger locality.

    \item \textbf{Internal edge fraction:}
    \[
    \text{internal\_frac}_m = \frac{|\{ (u,v)\in E : u,v \in \mathcal{L}_m \}|}{|\{ (u,v)\in E : u \in \mathcal{L}_m \ \text{or}\ v \in \mathcal{L}_m \}|}
    \]
    Fraction of edges touching the mask that remain inside it. Higher values indicate that the lightcone is self-contained.

    \item \textbf{Boundary ratio:}
    \[
    \text{boundary}_m = \frac{|\{ (u,v)\in E : (u \in \mathcal{L}_m, v \notin \mathcal{L}_m) \ \text{or vice versa} \}|}{|\{ (u,v)\in E : u \in \mathcal{L}_m \ \text{or}\ v \in \mathcal{L}_m \}|}
    \]
    Proportion of edges crossing from the lightcone to the rest of the circuit. Lower values indicate stronger isolation.

    \item \textbf{Pairwise Jaccard overlap:}
    \[
    J(m,n) = \frac{|\mathcal{L}_m \cap \mathcal{L}_n|}{|\mathcal{L}_m \cup \mathcal{L}_n|}
    \]
    Measures the overlap between lightcones of different qubits. Smaller values mean the masks are more distinct.

\end{itemize}

\begin{table}[ht]
\centering
\caption{Lightcone locality metrics for a training graphs with mean $N=10829$ nodes and $M=5$ measured qubits.}
\label{tab:lightcone_locality}
\begin{tabular}{lccc}
\hline
Qubit & Coverage & Internal frac & Boundary \\
\hline
$q0$ & 0.0147 & 0.0484 & 0.9516 \\
$q1$ & 0.0142 & 0.0469 & 0.9531 \\
$q2$ & 0.0142 & 0.0469 & 0.9531 \\
$q3$ & 0.0149 & 0.0495 & 0.9505 \\
$q4$ & 0.0099 & 0.0342 & 0.9658 \\
\hline
\multicolumn{4}{l}{\textbf{Global summary:}} \\
\multicolumn{4}{l}{Mean coverage = 0.0136 \quad (lower $\rightarrow$ more local)} \\
\multicolumn{4}{l}{Mean internal edge frac = 0.0452 \quad (higher $\rightarrow$ more isolated from global)} \\
\multicolumn{4}{l}{Mean boundary ratio = 0.9548 \quad (lower $\rightarrow$ more isolated from global)} \\
\multicolumn{4}{l}{Mean pairwise Jaccard = 0.0840 \quad (overlap across qubits)} \\
\hline
\end{tabular}
\end{table}

Table \ref{tab:lightcone_locality} clearly demonstrate the locality of the lightcone masks (very low average coverage scores) as well as their high relation with the global context (high boundary ratio and internal edge fractions), emphasizing the importance of using both lightcone local and global circuit-wide features. 
\section{Conclusion}

We presented QAGT-MLP; an attention-based graph neural network to be employed in practical and large quantum circuits error mitigation (QEM). QAGT-MLP encodes each quantum circuits as a graph in which different gates and qubit connectivity are translated into different nodes and and edges respectively. Using multi-head attention, QAGT-MLP extracts two sets of features from the input graph: 1) graph-wide global features and 2) local lightcone mask features. It then concatenates these features with circuit features and the noisy outputs of the circuit and feeds the resulting feature map into an MLP for final noise-mitigated value estimation. The experimental results on large-scale 100-qubit trotterized 1D TFIM circuits demonstrate the validity and applicability of the proposed QAGT-MLP on real-world practical QEM tasks. QAGT-MLP outperforms the state-of-the-art machine learning methods in both average error and its standard deviation.

\section{Acknowledgment}

This work was done during a summer internship in the University of Missouri's Quantum Innovation Center (Mizzou QIC). 

\bibliographystyle{unsrt}  
\bibliography{references}

\end{document}